\documentclass[twocolumn,superscriptaddress,showpacs,prl]{revtex4}
\usepackage{graphicx,epsfig,psfrag}

\newcommand{\be}{\begin{equation}}
\newcommand{\ee}{\end{equation}}
\newcommand{\al}{\alpha}

\newcommand{\bea}{\begin{eqnarray}}
\newcommand{\eea}{\end{eqnarray}}

\newcommand{\nn}{\nonumber}

\begin{document}

\noindent
TTP05-04\hfill 
\title{
\boldmath 
Two-Loop Electroweak Logarithms  
\unboldmath}
\author{Bernd Jantzen}
\affiliation{Bernd Feucht in previous publications}
  \affiliation{Institut f\"ur Theoretische Teilchenphysik,
    Universit\"at Karlsruhe, 76128 Karlsruhe, Germany}
\author{Johann H. K\"uhn}
\affiliation{Institut f\"ur Theoretische Teilchenphysik,
    Universit\"at Karlsruhe, 76128 Karlsruhe, Germany}
\author{Alexander A. Penin}
  \affiliation{Institut f\"ur Theoretische Teilchenphysik,
    Universit\"at Karlsruhe, 76128 Karlsruhe, Germany}
  \affiliation{Institute for Nuclear Research,
    Russian Academy of Sciences, 117312 Moscow, Russia}
\author{Vladimir A. Smirnov}
  \affiliation{Institute for Nuclear Physics,
    Moscow State University, 119992 Moscow, Russia}
\affiliation{II. Institut f{\"u}r Theoretische Physik,
  Universit{\"a}t Hamburg,  22761 Hamburg, Germany}


\begin{abstract}
  We present the complete analytical result for the two-loop
  logarithmically enhanced contributions to the high energy asymptotic
  behavior of the vector form factor and the four-fermion cross section
  in a spontaneously broken $SU(2)$ gauge model.  On the basis of this
  result we derive the dominant two-loop electroweak corrections to the
  neutral current four-fermion processes at high energies.
\end{abstract}

\maketitle

Recently a new wave of interest to the Sudakov asymptotic regime
\cite{Sud,Jac} has been triggered by the study of higher-order
corrections to electroweak processes at high energies
\cite{Kur,Bec1,Bec2,Fad,KPS,CCC,KMPS,HKK,FKM,FKPS}.  Experimental and
theoretical studies of electroweak interactions have traditionally
explored the range from very low energies, {\it e.g.} through parity
violation in atoms, up to energies comparable to the masses of the $W$-
and $Z$-bosons, {\it e.g.} at the LEP or the Tevatron.  The advent of
multi-TeV colliders like the LHC during the present decade or a future
linear electron-positron collider will give access to a completely new
energy domain. Once the characteristic energies $\sqrt{s}$ are far
larger than the masses of the $W$- and $Z$-bosons, $M_{W,Z}$, exclusive
reactions like electron-positron (or quark-antiquark) annihilation into
a pair of fermions or gauge bosons will receive virtual corrections
enhanced by powers of the large {\it electroweak} logarithm
$\ln\bigl({s/ M_{W,Z}^2}\bigr)$.  The leading double-logarithmic
corrections may well amount to ten or even twenty percent in one-loop
approximation and reach  several percent in two-loop approximation.
Moreover, in the TeV region, the subleading logarithms turn out to be
equally important \cite{KPS,KMPS}. One percent accuracy of the
theoretical estimates for the cross sections necessary for the search of
new physics beyond the standard model can be guaranteed only by
including {\it all} the logarithmic two-loop corrections.

The direct calculation of the two-loop electroweak corrections is an
extremely challenging theoretical problem at the limit of available
computational techniques even in the high energy limit.  However, the
asymptotic high energy behavior of the amplitudes is governed by
evolution equations which turn out to be a powerful tool in the analysis
of the logarithmic corrections.  In Ref.~\cite{Fad} the leading
logarithmic (LL) electroweak corrections have been obtained to all
orders of perturbative expansion within the infrared evolution equation
approach.  This analysis has been extended to the NLL and N$^2$LL
approximation \footnote{N$^m$LL stands for the corrections of the form
  $\al^{n}\ln^{2n-m}(s)$ for an arbitrary $n$.} in Refs.~\cite{KPS,KMPS}
by combining the hard and infrared evolution equations. Starting with
the N$^3$LL approximation the corrections become sensitive to fine
details of the gauge boson mass generation and the analysis is
complicated by the presence of the mass gap and mixing in the gauge
sector.  In Ref.~\cite{FKPS} the general matching procedure has been
formulated which relates theories with and without mass gap, thus
setting the stage for the calculation of the logarithmically enhanced
two-loop corrections to electroweak processes.  In this Letter the
analysis of Ref.~\cite{FKPS} will be completed. We first present
explicit analytical results for the two-loop logarithmic corrections to
the vector form factor and four-fermion cross section in the
spontaneously broken $SU(2)$ model with the gauge and Higgs bosons of
the same mass $M$ and six doublets of left-handed massless fermions
inspired by the standard model.  Then we proceed along the line of
Ref.~\cite{FKPS} and derive the numerical results for the dominant
two-loop electroweak corrections to the cross sections of the neutral
current four-fermion processes in the full $SU_L(2)\times U(1)$ theory
with light fermions. We neglect the fermion mass effects which can be
important for the top and bottom quark production.

The vector form factor ${\cal F}$ determines the fermion scattering
amplitude in an external Abelian field.  It plays a special role since
it is the simplest quantity which includes the complete information
about the universal {\it collinear} logarithms  directly
applicable to any process with an arbitrary number of fermions.  Let us
write the perturbative expansion for the form factor as ${\cal
  F}=\sum_n\left(\al\over 4\pi\right)^nf^{(n)}$, where $f^{(0)}=1$
corresponds to the Born approximation and the coupling constant $\al$ is
renormalized at the scale $M$ according to $\overline{\rm MS}$
prescription.  In the $SU(2)$ model the one-loop coefficient $f^{(1)}$
in the Sudakov limit $M/Q\to 0$ can easily be obtained from the known
$U(1)$ result (see {\it e.g.} \cite{FKPS}) by multiplying with the group
factor $3/4$.  For the two-loop logarithmic contribution of the virtual
gauge and Higgs bosons we find by explicit calculation
\bea
\lefteqn{f^{(2)}=
{9\over 32}\bar{\cal L}^4
+{5\over 48}\bar{\cal L}^3-\left({691\over 48}-{7\over 8}\pi^2\right)\bar{\cal L}^2
+\left({167\over 4}\right.}
\nn\\
&&
\left.-{11\over 24}\pi^2
-{61\over 2}\zeta_3+{15\over 4}\sqrt{3}\pi
+{13\over 2}\sqrt{3}{\rm Cl}_2\left({\pi\over 3}\right)\right)\bar{\cal L}
\nn\\
&&
+{\cal O}(\bar{\cal L}^0)\,.
\label{2loopf}
\eea where $\bar{\cal L}=\ln\left({Q^2/M^2}\right)$, $Q$ is the Euclidean
momentum transfer, all power-suppressed terms are neglected,
$\zeta_3=1.202057\ldots$ and ${\rm Cl}_2(\pi/3)=1.014942\ldots$ are the
values of the Riemann's zeta-function and the Clausen function,
respectively.  In Eq.~(\ref{2loopf}) we do not include the contribution
due to the virtual fermion loop computed in \cite{FKM}.  The Abelian
contribution to Eq.~(\ref{2loopf}) has been evaluated in
Ref.~\cite{FKPS}.  For the calculation of the leading power behavior of
the two-loop on-shell vertex diagrams with two massive propagators in
the Sudakov limit we used the expansion by regions approach
\cite{BenSmi} (for the application to the Sudakov form factor see also
\cite{KPS}).  The method is based on the separation of the
contributions of the dynamical modes characteristic for the Sudakov
limit \cite{Ste} in dimensional regularization.  Our result for the
hard modes agrees with the dimensionally regularized
massless result of Ref.~\cite{KraLam}.  The result for the coefficients
of the quartic, cubic and quadratic logarithms in Eq.~(\ref{2loopf}) is
in full agreement with the predictions of the evolution equation
approach \cite{KMPS}.  In particular, they are not sensitive to details
of the gauge boson mass generation.  This is not true for the
coefficient of the linear-logarithmic term which depends {\it e.g.} on
the Higgs boson mass.  For example, in the (hypothetical) case of a
light Higgs boson with mass $M_H\ll M$ the coefficient of the linear
logarithm in Eq.~(\ref{2loopf}) becomes \be {333\over 8}-{11\over
  48}\pi^2 -{61\over 2}\zeta_3+{33\over 8}\sqrt{3}\pi +{21\over
  4}\sqrt{3}{\rm Cl}_2\left({\pi\over 3}\right)\,.
\label{zeromh}
\ee 
Let us now consider the four-fermion process $f\bar f\to f'\bar f'$.  We
define the perturbative expansion for the corresponding normalized total
cross section as follows: ${\cal R}\equiv \sigma/\sigma_{\rm Born}
=\sum_n\left(\al\over 4\pi\right)^nr^{(n)}$, $r^{(0)}=1$, where the
coupling constant in the Born cross section is renormalized at the scale
$\sqrt{s}$ while the series in $\al$ is renormalized at the scale $M$.
The one-loop coefficient $r^{(1)}$ in the Sudakov limit can be found in
Ref.~\cite{KMPS}.  The four-fermion amplitude can be decomposed into
(the square of) the form factor and a {\it reduced} amplitude
\cite{KPS,KMPS}.  The latter carries all the Lorentz and isospin indices
and does not contain collinear logarithms.  The logarithmic corrections
to the reduced amplitude are obtained by solving a renormalization group
like equation \cite{Sen}.  The corresponding two-loop anomalous
dimensions can be extracted from the existing massless QCD calculations
\cite{AGOT,Glo,FreBer} (see \cite{KMPS,SteTej}), or obtained within the Wilson line
approach \cite{ADS}.  By combining Eq.~(\ref{2loopf}) with the result
for the reduced amplitude and integrating the cross section over the
production angle we obtain the two-loop logarithmic contribution
\bea
\lefteqn{r_+^{(2)}={9\over 2}{\cal L}^4
-{449\over 6}{\cal L}^3
+\left({4855\over 18}+{37\over 3}\pi^2\right){\cal L}^2
+\left({48049\over 216}\right.}
\nn\\
&&
\left.-{1679\over 18}\pi^2-122\zeta_3+15\sqrt{3}\pi
+26\sqrt{3}{\rm Cl}_2\left({\pi\over 3}\right)\right){\cal L}
\nn\\
&&
+{\cal O}({\cal L}^0)\,,
\label{2looprp}
\eea
and
\bea
\lefteqn{r_-^{(2)}={9\over 2}{\cal L}^4
-{125\over 6}{\cal L}^3
-\left({799\over 9}-{37\over 3}\pi^2\right){\cal L}^2
+\left({38005\over 216}\right.}
\nn\\
&&
\left.-{383\over 18}\pi^2-122\zeta_3+15\sqrt{3}\pi
+26\sqrt{3}{\rm Cl_2}\left({\pi\over 3}\right)\right)
{\cal L}
\nn\\
&&+{\cal O}({\cal L}^0)\,,
\label{2looprm}
\eea 
for the initial and final state fermions of the same or opposite
isospin, respectively.  In Eqs.~(\ref{2looprp},~\ref{2looprm}) the
virtual fermion loop contribution is included and ${\cal L}=\ln(s/M^2)$
is real in the physical region of positive $s=-Q^2$.  The
coefficients of the quartic, cubic and quadratic logarithms in
Eqs.~(\ref{2looprp},~\ref{2looprm}) are already given in Ref.~\cite{KMPS},
the linear-logarithmic term is new.

The main distinction of the analysis in the standard electroweak model
with the spontaneously broken $SU_L(2)\times U(1)$ gauge group from the
pure $SU_L(2)$ case is the presence of the massless photon which results
in infrared divergences in fully exclusive cross sections.  The
divergences are cancelled in cross sections which are inclusive with
respect to the soft photon bremsstrahlung. Besides the electroweak
logarithms the inclusive cross sections get logarithmic corrections of
the form $\ln(s/\varepsilon_{cut}^2)$ and $\ln(s/m^2)$ where $m$ is an
initial or final state fermion mass and $\varepsilon_{cut}$ is the soft
photon energy cut. In the case $m,~\varepsilon_{cut}\ll M_{W,Z}$ these
logarithms are of pure QED nature and are known to factorize.  Note that
the two-loop pure QED corrections to the four-fermion cross section are
known even beyond the logarithmic approximation (see \cite{Pen} and
references therein).  Within the evolution equation approach \cite{Fad}
it has been found \cite{KMPS} that the electroweak and QED logarithms up to the
N$^2$LL approximation can be disentangled by means of the following
two-step procedure: (i) the corrections are evaluated using the fields
of the unbroken symmetry phase with all the gauge bosons of the same mass
$M\approx M_{Z,W}$; (ii) the QED contribution with an auxiliary photon
mass $M$ is factorized leaving the pure electroweak logarithms. This
reduces the calculation of the two-loop electroweak logarithms up to the
quadratic term to a problem with a single mass parameter.  Then the
effect of the $Z-W$ boson mass splitting can systematically be taken
into account within an expansion around the equal mass approximation
\cite{FKPS}.  In general the above two-step procedure is not valid in
the N$^3$LL approximation which is sensitive to fine details of the
gauge boson mass generation.  For the exact calculation of the
coefficient of the two-loop linear-logarithmic term one has to use the
true mass eigenstates of the standard model.  The evaluation of the
corrections in this case becomes a very complicated multiscale problem.
The analysis, however, is drastically simplified in a model with a Higgs
boson of zero hypercharge.  In this model the mixing is absent and the
above two-step procedure can be applied to disentangle all the two-loop
logarithms of the $SU_L(2)$ gauge boson mass from the infrared
logarithms associated with the massless hypercharge gauge boson
\cite{FKPS}.  With the result for the $SU_L(2)$ model presented in this
Letter at hand we are able to complete the analysis of the two-loop
logarithmic corrections in the simplified model.  In the standard model
the mixing of the gauge bosons results in a linear-logarithmic
contribution which is not accounted for in this approximation.  It is,
however, suppressed by a small factor $\sin^2{\theta_W}\approx 0.2$,
with $\theta_W$ being the Weinberg angle.  Therefore, the approximation
gives an estimate of the coefficient in front of the linear electroweak
logarithm with $20\%$ accuracy.

\begin{figure}[b]
\epsfig{figure=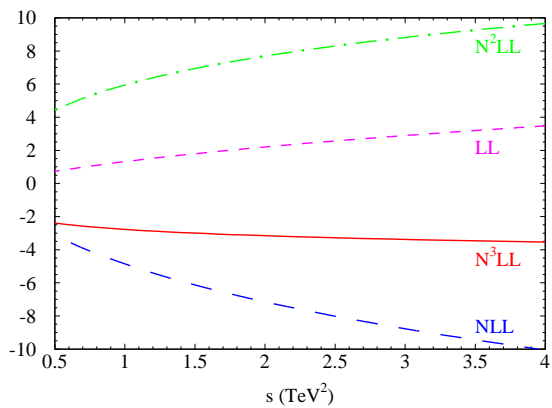,height=6cm}
\caption{\label{fig1} 
  \small The LL (short-dashed line), NLL (long-dashed line), NNLL
  (dot-dashed line) and N$^3$LL (solid line) two-loop electroweak
  corrections to ${\cal R}_{lq}$ in percent.}
\end{figure}

\begin{figure}[b]
\epsfig{figure=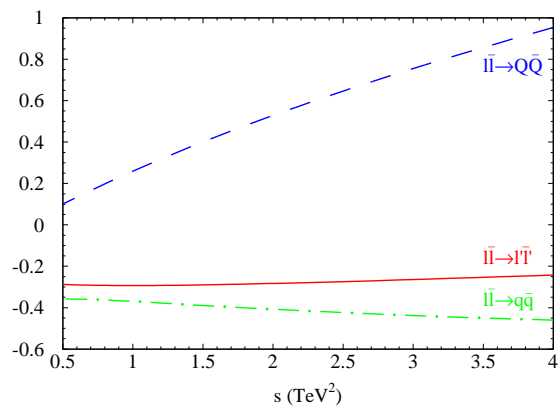,height=6cm}
\caption{\label{fig2}  The total electroweak logarithmic  two-loop corrections
  to ${\cal R}_{lQ}$ (dashed line), ${\cal R}_{lq}$ (dot-dashed line)
  and ${\cal R}_{ll'}$ (solid line) in percent.}
\end{figure}

Let ${\cal R}_{ff'}\equiv \sigma/\sigma_{em}$ be the normalized total
cross section of the $f\bar f$ annihilation into a $f'\bar f'$ pair.
Here $\sigma_{em}$ stands for the cross section which incorporates the
pure QED radiative corrections and is free of the electroweak
logarithms.  It is convenient to normalize $\sigma_{em}$ so that the
virtual QED corrections vanish at $m=0,~s=\lambda^2$, where $\lambda$ is
the auxiliary photon mass, and to use the electroweak coupling constants
renormalized at the scale $\sqrt{s}$ in the Born approximation
\cite{KMPS}.  In the standard electroweak model the perturbative
expansion involves two parameters: the $SU_L(2)$ coupling constant $\al$
and the $U(1)$ hypercharge coupling constant $\al'$. For convenience we
eliminate the latter by means of the relation $\al'=\tan^2\theta_W\,\al$
and define the one-parameter series for the cross section ${\cal
  R}_{ff'} =\sum_n\left(\al\over 4\pi\right)^nr_{ff'}^{(n)}$,
$r_{ff'}^{(0)}=1$, in terms of the $\overline{\rm MS}$ $SU_L(2)$ coupling
renormalized at the scale of the gauge boson mass.  The complete
one-loop result for the cross section is known exactly (see {\it e.g.}
Ref.~\cite{BHM} and references therein).  For the two-loop logarithmic
corrections to the phenomenologically interesting processes we obtain
the following numerical approximation:
\bea
r^{(2)}_{lQ}
&=&1.93\,{\cal L}^4-11.28\,{\cal L}^3+33.79\,{\cal L}^2-60.87 \,{\cal L} \,,
\nn \\
r^{(2)}_{lq}
&=&2.79\,{\cal L}^4-51.98\,{\cal L}^3+321.20\,{\cal L}^2-757.35 \,{\cal L}\,,
\nn\\
r^{(2)}_{Qq}
&=&3.53\,{\cal L}^4-20.39\,{\cal L}^3+65.20\,{\cal L}^2-91.92 \,{\cal L}\,,
\nn\\
r^{(2)}_{ll'}
&=&1.42\,{\cal L}^4-20.33\,{\cal L}^3+112.57\,{\cal L}^2-312.90 \,{\cal L}\,,
\nn\\
r^{(2)}_{QQ'}
&=&2.67\,{\cal L}^4-46.64\,{\cal L}^3+278.94\,{\cal L}^2-666.05 \,{\cal L}\,,
\nn\\
r^{(2)}_{qq'}
&=&4.20\,{\cal L}^4-71.87\,{\cal L}^3+423.61\,{\cal L}^2-919.35 \,{\cal L}\,,
\nn\\&&
\label{finres}
\eea 
where ${\cal L}=\ln\left({s/M_W^2}\right)$, $l$ stands for a charged
lepton, $Q$ and $q$ stand for the $u,~c,~t$ and $d,~s,~b$ quarks,
respectively.  Note that the result is symmetric under exchange of the
initial and final state fermions and can easily be generalized to $f\bar
f \to f\bar f$ processes by including the $t$ channel contribution which
goes beyond the scope of this Letter.  In Eq.~(\ref{finres}) we use the
value $\sin^2{\theta_W}=0.231$ corresponding to the $\overline{\rm MS}$
coupling constants renormalized at the scale $M_Z$.  The coefficients of
the cubic and quadratic logarithms in the two-loop corrections to the
cross sections of $e^+e^-$ annihilation have been computed in
Refs.~\cite{KPS,KMPS} neglecting the $W-Z$ boson mass difference
\footnote{Throughout Sect~4.  of Ref.~\cite{KMPS} the terms with the
  factor $(aN_g+b)t_W^2$, where $a$ and $b$ stand for some constants,
  should be multiplied by an extra $t_W^2$, and the terms with the
  factor $N_gs_W^2$ should be multiplied by $s_W^2$. This results in a
  small change of the numerical estimates.}.  
In Eq.~(\ref{finres}) we included the leading correction in the mass
difference $1-M_W/M_Z$ to these coefficients.  The coefficient of the
linear logarithm is computed in the approximation described above.

For the case of $e^+e^-$ annihilation the size of the corrections is
shown in Figs.~\ref{fig1},~\ref{fig2}.  In Fig.~\ref{fig1} the values of
different two-loop logarithmic contributions to ${\cal R}_{lq}$ are
plotted separately as functions of $s$ for $\al = 3.38\cdot 10^{-2}$.
The two-loop logarithmic terms have a sign-alternating structure
resulting in significant cancellations.  Although the individual
logarithmic contributions can be as large as $10\%$, their sum does not
exceed $1\%$ at energies below 2~TeV for all the cross sections (see
Fig.~\ref{fig2}).  In the region of a few TeV the corrections do not
reach the double-logarithmic asymptotics. The quartic, cubic and
quadratic logarithms are comparable in magnitude.  The
linear-logarithmic term is a few times smaller than the quadratic
logarithm which is in agreement with the prediction of Ref.~\cite{KMPS} for
the structure of the two-loop corrections and justifies neglecting the
nonlogarithmic contribution.  Still, the linear-logarithmic
contribution amounts to a few percent and must be included to reduce
the theoretical uncertainty below $1\%$.

Let us discuss the accuracy of our result.  On the basis of the explicit
evaluation of the light fermion/scalar \cite{FKM} and the Abelian
contribution \cite{FKPS} we estimate the uncalculated two-loop
nonlogarithmic term to few permill.  For $\sqrt{s}>500$~GeV the
power-suppressed terms do not exceed a permill in magnitude as well
\cite{FKM}.  The leading effect of the $W-Z$ mass splitting results in a
variation of the coefficients of the two-loop cubic and quadratic
logarithms of at most $5\%$. Thus the expansion in the $W-Z$ mass
difference converges well for these coefficients and the leading
correction term taken into account in our evaluation is sufficient for a
permill accuracy of the cross sections.  Neglecting the gauge boson
mixing effects, which are suppressed by a factor of $\sin^2{\theta_W}$,
induces an error of $20\%$ in the coefficient of the two-loop single
logarithm.  Neglecting the difference between the Higgs and gauge boson
masses leads to a variation of the linear logarithmic coefficient of at
most $5\%$ since the scalar boson contribution is relatively small.  The
same is true for the uncertainty due to the top quark mass effect on the
$t\bar t$ virtual pair contribution.  Hence for the production of light
fermions our formulae are supposed to approximate the exact coefficients
of the two-loop linear logarithms with approximately $20\%$ accuracy
leading to a few permill uncertainty in the cross sections. By adding up
the errors from different sources in quadrature we find the total
uncertainty of the cross section to be from a few permill up to one
percent, depending on the process.  This result should be sufficient for
all practical applications to collider physics.  The only essential
deviation of the exact two-loop logarithmic contributions from our
result is relevant for the production of the third generation quarks and
is due to the large top quark Yukawa coupling.  The corresponding
corrections are known to NLL approximation and can numerically be as
important as the generic non-Yukawa ones \cite{Bec2}.

To conclude, we have derived the analytical result for the two-loop
logarithmic corrections to the vector form factor and four-fermion cross
section in the spontaneously broken $SU_L(2)$ model.  We have also
obtained the dominant two-loop electroweak corrections to neutral
current four-fermion processes, which are crucial for the high-precision
physics at the LHC and the next generation of linear colliders.

\begin{acknowledgments}
  We thank the authors of Refs.~\cite{Glo,SteTej,ADS} for very usefull
  communications.  The work of J.H.K and A.A.P.  was supported in part
  by BMBF Grant No.\ 05HT4VKA/3 and SFB/TR9. The work of V.A.S. was
  supported in part by the Russian Foundation for Basic Research through
  project 05-02-17645 and DFG Grant No. Ha 202/110-1.
\end{acknowledgments}

\end{document}